\documentclass{elsart}
\def\ohalf{{\textstyle{1\over 2}}}

\def\thalf{{\textstyle{3\over 2}}}

\def\P{{\sf P}}

\def\MD{M_{_\Delta}}

\def\MN{M_{_N}}

\newcommand{\beq}{\begin{equation}}
\newcommand{\be}{\begin{equation}}
\newcommand{\eeq}{\end{equation}}
\newcommand{\ee}{\end{equation}}
\newcommand{\beqa}{\begin{eqnarray}}
\newcommand{\bea}{\begin{eqnarray}}
\newcommand{\eeqa}{\end{eqnarray}}
\newcommand{\eea}{\end{eqnarray}}
\newcommand{\bra}[1]{\langle {#1} |}                        
\newcommand{\ket}[1]{| {#1} \rangle}
\newcommand{\expect}[2]{\langle #1 | #2 \rangle}

\usepackage{graphicx}
\usepackage{bm,hyperref,color}
\usepackage{epsfig}

\begin{document}
\begin{frontmatter}
\title{D-state configurations in the electromagnetic form
factors of the nucleon and the ${\mathbf\Delta}{\bf (1232)}$ resonance}

\author[pit,hel]{B. Juli\'a-D\'{\i}az}
\author[hel]{D.O. Riska}

\address[pit]{Department of Physics and Astronomy, 
University of Pittsburgh, PA 15260, USA}

\address[hel]{Helsinki Institute of Physics and Department of Physical
Sciences, P.O.Box 64, 00014 University of Helsinki, Finland}

\date{}

\begin{abstract}
The $\Delta-N$ electromagnetic transition form 
factors are calculated in the Poincar\'e covariant 
quark model in three forms of relativistic kinematics.
Addition of $D-$state components to pure $S-$state model 
wave functions, chosen so as to reproduce the empirical 
elastic electromagnetic nucleon form factors with single
constituent currents, brings the calculated $R_{EM}$ 
ratio for the $\Delta(1232)\rightarrow N\gamma$ transition 
closer to the empirical values in instant and point 
form kinematics. The calculated $R_{SM}$ ratio is 
insensitive to the $D-$state component. In front form 
kinematics the substantial violation of the angular 
condition for the spin 3/2 resonance transition
amplitude in the impulse approximation prevents a unique 
determination of $R_{EM}$ and $R_{SM}$, both of which 
are very sensitive to $D-$state components. In no form 
of kinematics do $D-$state deformations of the rest frame 
baryon wave functions alone suffice for a description
of the empirical values of these ratios.
\end{abstract}
\end{frontmatter}

\section{Introduction}

Deformation of the nucleon wave function from 
spherical symmetry in the rest frame, and its possible empirical 
manifestations have drawn considerable experimental 
and theoretical attention~\cite{gmiller1,ralston,bern,mainz}. 
The experimental finding of nonvanishing $E2/M1$ and 
$C2/M1$ ratios in the electromagnetic decay of the 
$\Delta(1232)$ suggests the presence either of spatial 
$D-$state components in the wave functions or of 
quark-antiquark configurations 
(``meson cloud effects~\cite{satolee,satolee2}'').

The elastic form factors of the proton and the magnetic 
form factor of the neutron may be described in the 
constituent quark model, with point-like quarks and 
spherically symmetric $S-$state rest frame wave functions 
in all the three forms of relativistic kinematics outlined 
by Dirac~\cite{Dirac}. With an additional small ($\sim$ 1\%) 
admixture of a mixed symmetry S-state component the electric 
form factor of the neutron may be described as well~\cite{bruno}.
The ranges of the wave functions do however differ 
significantly between the different forms of kinematics, 
as do the high-$Q^2$ behaviors of the form factors.

In the covariant quark model both the electric form 
factor of the nucleon and the $E2/M1$ and $C2/M1$ ratios 
in $\Delta\rightarrow N\gamma$ decay are non-zero even in 
the case of spatially symmetric $S-$state three-quark wave functions 
in the rest frame~\cite{Weber}. This is a consequence of 
the kinematic boosts and the Wigner rotations
(or Melosh rotations in the case of front form kinematics), 
which are absent in the non-relativistic quark model. The key 
question then becomes to what extent admixtures of components 
with higher angular momenta or of quark-antiquark components
may be required in the baryon wave functions for a 
satisfactory description of the empirical observables.

It is shown here that the effect of adding $D-$state 
components to the rest frame wave functions of the proton 
and the $\Delta(1232)$ resonance is significant for 
the $E2/M1$ ratio $R_{EM}$, but very small for the 
$C2/M1$ ratio $R_{SM}$ in the case of instant and point 
form kinematics. It is found that with single quark 
currents the notable empirical (negative) peak in $R_{EM}$ 
below 0.5 GeV$^2$ cannot be described by $D-$state 
deformations alone. It is also found that the covariant 
quark model yields small positive values for $R_{SM}$, 
in disagreement with the empirically found negative values.

In the case of the $\Delta(1232)\rightarrow N\gamma$ transition, 
front form kinematics does not yield unique results for 
the $E2/M1$ and $C2/M1$ ratios, because of the violation 
of the linear relation between more than three different
spin amplitudes in the front form 
description of form factors of states with spin 3/2 
and single constituent current 
operators~\cite{Coester92,simula,simula2}.
The calculated magnetic $\Delta(1232)\rightarrow N\gamma$ 
transition form factor is, however, insensitive to this 
problem, and is qualitatively similar to that obtained
in instant and front form kinematics. 

The article is organized as follows. In Section~\ref{sec:dnt} 
the calculation of the transition form factors from the 
matrix elements of the electromagnetic current of the 
constituent quarks is outlined for all the three forms 
of kinematics. In Section~\ref{sec:wf} the explicit baryon 
wave function model that includes $D-$state components is 
described. In Section~\ref{sec:ND} the matrix elements 
required for the calculation of the $\Delta(1232)\rightarrow
N\gamma$ 
transition are given explicitly. In Section~\ref{sec:eds} 
the effects of the $D-$state components are studied in 
detail. A concluding summary is provided in the last section.

\section{${\mathbf\Delta}$(1232)-N${\mathbf\gamma}$
transition form factors}
\label{sec:dnt}

The electromagnetic elastic and transition form factors 
of the baryons are linear combinations of matrix elements 
of the electromagnetic current. In instant and point form 
kinematics the appropriate matrix elements for the 
$\Delta(1232)\rightarrow N\gamma$ transition form factors are 
matrix elements of the $I_1$ and $I_0$ components of 
the current operator. 

The general expression for the transition current operator:
\beq
\bra{\Delta} {\mathcal I}^\mu(0) \ket{ N} = 
e \bar \Psi_\nu(p^*) \Gamma^{\nu \mu} \Psi(p) \, ,
\eeq
has the conventional decomposition~\cite{bruno}:
\beq
\Gamma^{\nu \mu} = \sum_i G_i(Q^2)\, {\mathcal K}^{\nu\mu}_i\, .
\eeq
The operators ${\mathcal K}^{\nu\mu}_i$ are defined as:
\beqa
&&{\mathcal K}^{\nu\mu}_1={Q^\nu\gamma^\mu-(\gamma\cdot
Q)g^{\nu\mu}\over \sqrt{Q^2}}\sqrt{M^* M}\gamma_5\, ,
\nonumber \\
&&{\mathcal K}^{\nu\mu}_2={Q^\mu P^\nu-(P\cdot
Q)g^{\nu\mu}\over \sqrt{Q^2}}\gamma_5\, ,
\nonumber \\
&&{\mathcal K}^{\nu\mu}_3={Q^\nu Q^\mu-Q^2
g^{\nu\mu}\over \sqrt{Q^2}}M^*\gamma_5\, .
\label{ks}
\eeqa
Here $M$ and $M^*$ are the masses of the nucleon and 
$\Delta(1232)$. The relation between the three form 
factors $G_1, G_2$ and $G_3$ and the corresponding
electric, magnetic and Coulomb form factors are~\cite{Scadron}:
\beqa
 G_E^*&=&{M\over 3(M^*+M)} \left[{M^{*2}-M^2-Q^2\over M^*}
{\sqrt{M^* M} \over Q} G_1\right. \nonumber\\
&+&\left.{M^{*2}-M^2\over Q}G_2-2M^*G_3 \right]\, ,\nonumber \\
 G_M^*&=&{M\over 3(M^*+M)}\left[{(3M^*+M)(M^*+M)+Q^2\over M^*}
{\sqrt{M^* M} \over Q}G_1 \right.\nonumber \\
   &+& \left.{M^{*2}-M^2\over Q} G_2-2 M^* G_3\right]\, ,
\nonumber \\
  G_C^*&=&{2M\over 3(M^*+M)}\left[{ 2 M^*\sqrt{M^* M} \over Q} G_1 +
{3M^{*2}+M^2+Q^2 \over 2 Q} G_2 \right.\nonumber \\
&+& \left.{M^{*2}-M^2-Q^2\over Q^2}  M^*G_3\right]  \, .
\label{oursEMC}
\eeqa
In instant and point form kinematics the relation 
between the different spin state matrix elements of 
the electromagnetic current and the form factors, 
$G_j$, is finally:
\begin{eqnarray}
I_{\thalf,\ohalf}^1
&=&\left[ {M^* +M\over \sqrt{ Q^2}} G_1 + 
{M^{*2}-M^2\over 2\sqrt{Q^2 M M^*}}G_2
-{\sqrt{M^*\over M} }G_3\right]
  {Q_3\over 2 \sqrt{E(M+E)}} \, ,
\nonumber\\
I_{\ohalf,-\ohalf}^1
&=&-{\sqrt{3}\over 6}\left[{M^*+M\over \sqrt{ Q^2}}G_1
+ {M^{*2}-M^2\over 2\sqrt{ Q^2 M M^*}}G_2-
{\sqrt{M^*\over M}}G_3\right]
  {Q_3\over \sqrt{E(M+E)}}  \nonumber\\
&+&{\sqrt{3}\over 3}
{Q_3\over \sqrt{Q^2}}{M+E\over \sqrt{E(M+E)}}G_1\, ,
\label{Gs}  \nonumber \\
I_{\ohalf,\ohalf}^0
&=&- {\sqrt{3}\over 3}
\left[{Q_3\over \sqrt{Q^2}}G_1
+{Q_3\over \sqrt{Q^2M M^*}}{E+M^*\over 2} G_2\right. \nonumber\\ 
&+&\left. {Q_3 Q_0 \over Q^2}\sqrt{M^*\over M} G_3 \right] 
 {Q_3 \over \sqrt{ E (M+E)} } \, . 
\end{eqnarray}
Here the 4-momentum transfer is taken as 
$Q=\{Q^0,0,0,Q_3\}$, 
with $I_{{j_\Delta},{j_N}}^m=\langle j_\Delta, 
P_\Delta\vert I_m(0)\vert P_N,j_N \rangle$,
and
\beq
Q^0 = -{P^* \cdot Q \over M^*} =  {M^{*2} - M^2 -Q^2\over 2 M^*}\,, 
\qquad  Q_3=\sqrt{Q^2+Q^{0\,2}} \, .
\eeq

The $E2/M1$ and $C2/M1$ ratios for the $\Delta-N$ 
transition are defined 
as\footnote{The definition of $R_{SM}$ is that of 
Ref.~\cite{satolee2}, the overall sign of which
disagrees with that in Ref.~\cite{simula}.}:
\beqa
&&R_{EM}\equiv {E_2 \over M_1} \equiv - {  G_E^* \over G_M^*}\, , 
\nonumber \\
&& R_{SM}\equiv{C_2 \over M_1} \equiv  {|\vec q|\over 2 M^*}
 {  G_C^* \over G_M^*}\,.
\eeqa
Here $2M^* |\vec q|= 
\left([Q^2 + (M^*-M)^2][Q^2 + (M^*+M)^2]\right)^{1/2}$.

In the application of front form kinematics it has been
conventional to adopt a reference frame in which $Q^+=0$. 
In the case of elastic form factors the relation between 
the initial and final state is kinematic in this 
frame~\footnote{These relations together with 
Eq.~(\ref{oursEMC}) are equivalent to the definition 
of Eq.(3) of~\cite{simula} together with the form factors 
of Ref.~\cite{devenish}.}. 
The relation between the invariant form factors $G_j$ and 
the current matrix elements in this frame is 
\beqa
I_{\thalf,\ohalf}&=& {1\over \sqrt{2}} 
\left[G_1+ {M_{_\Delta} -M \over 
 \sqrt{4 MM_{_\Delta}}}G_2 \right] \, ,
\nonumber \\
I_{\ohalf,\ohalf}&=& -{1\over \sqrt{6}} \left[ 
Q {G_1 \over M_{_\Delta}} 
+Q {2M_{_\Delta}-M \over M_{_\Delta} \sqrt{ 4 MM_{_\Delta}}}G_2
-{M_{_\Delta}-M \over \sqrt{MM_{_\Delta}}} G_3
\right] \, ,
\nonumber \\
I_{\ohalf,-\ohalf}&=&-{1\over \sqrt{6}} \left[ 
-{M\over M_{_\Delta}} G_1 
+ { M_{_\Delta}(M_{_\Delta}-M) -Q^2 \over  M_{_\Delta}
 \sqrt{4MM_{_\Delta}}} G_2
+{Q\over \sqrt{MM_{_\Delta}}} \,G_3 
\right] \, ,
\nonumber \\
I_{\thalf,-\ohalf}&=& -{Q\over 2\sqrt{2MM_{_\Delta}}} G_2\, .
\label{Qplus0}
\eeqa
Here $I_{\nu_1,\nu_2}$ is defined as: 
\beq
I_{\nu_1,\nu_2}=\bra{\Delta,\nu_1} {\mathcal I}^+ \ket{N,\nu_2} \, ,
\eeq
where ${\mathcal I}^+ = n\cdot {\mathcal I}$ and
$n$ is the null-vector: $n=\{-1,0,0,1\}$.

These four spin amplitudes are linear combinations 
of only three invariant transition form factors and 
therefore they must are linearly dependent. This linear relation 
is broken by the sum of matrix elements of single quark currents 
in front form kinematics, and as a consequence, the 
calculated transition form factors depend on which set 
of three spin amplitudes are employed to calculate the 
transition form factors~\cite{simula,simula2}. This prevents 
definite predictions in front form kinematics for form 
factors of states with spin larger than 1/2 when 
only single quark currents are employed.

In the case of the transition form factors the $Q^+ = 0$ 
frame is not required by the kinematics, and has the 
disadvantage of requiring analytic extrapolation to the 
region of timelike momenta. For transition form factors 
it may be more appropriate to adopt a frame in which 
$Q^+ \neq 0$ and which allows extrapolation to timelike 
momenta with real kinematic coefficients~\cite{bruno2}. 

This frame may be defined as in Ref.~\cite{bruno2}:
\beq
v_f=\{\sqrt{1+\eta},\sqrt{\eta},0,0\}\, ,
 \qquad v_a=\{\sqrt{1+\eta},-\sqrt{\eta},0,0\}\, , 
\eeq
so that
\beqa
&&Q^+=(M_f-M_a)\sqrt{1+\eta} \, ,
\qquad Q_\perp= (M_f+M_a)\sqrt{\eta} \, , \nonumber \\
&& M_a p_f^+ = M_f p_a^+\, ,
 \qquad  M_a p_{f\perp} + M_f p_{a\perp} = 0\, .
\eeqa

In this frame the relation between the invariant 
form factors $G_j$ and the current matrix elements 
are:
\beqa
I_{\thalf,\ohalf}&=& - {\sqrt{2}  
\MN \sqrt{\eta} \over Q} \; G_1 \, ,
 \nonumber \\
I_{\ohalf,\ohalf}&=&\sqrt{\MN\over \MD} {\eta \over \sqrt{6}} \left[
4 \sqrt{\MD \MN \over Q^2 } \; G_1
+
{\MD+ \MN \over Q} \;G_2 \right. \nonumber\\
&+& \left.
{ 2 \MD (\MD - \MN )\over Q^2} \; G_3 \right]\, ,
 \nonumber \\
I_{\ohalf,-\ohalf}&=&
-{\sqrt{\eta} \over \sqrt{6}} 
\left[
2 {\MD\over Q} \;  G_1 
+
{ \MD^2-\MN^2+2 \eta \MN (\MD+ \MN ) \over Q \sqrt{\MN\MD}}\;  G_2
\right.\nonumber \\
&+&\left.
2 \sqrt{\MD\over \MN} {2 \eta \MN (\MD-\MN )- Q^2 \over Q^2}\;  G_3 
\right]\, ,
\nonumber \\
I_{\thalf,-\ohalf}&=&
\sqrt{\MN\over \MD} {\eta \over \sqrt{2}}  
\left[  
{\MD+\MN\over Q}\;  G_2
+ 
{2\MD(\MD-\MN) \over Q^2} \;   G_3 
\right]\, .
\label{Qplusne0}
\eeqa

In this case the last one of 
these spin amplitudes is a linear combination
of the first two.
This ``angular condition'' is broken by the 
matrix elements of the sum of single quark current 
operators in front form kinematics, and therefore 
the calculated transition form factors also in this 
case depend on which set of three amplitudes are used 
to determine the three invariant transition form factors.

\section{The nucleon and ${\mathbf\Delta}$(1232) wave functions}
\label{sec:wf}

The rest frame wave function of the nucleon shall be 
taken to be a combination of $S-$ and $D-$state 
components as:
\beq
\phi_N = a_N \phi_S + b_N \phi_D\, , 
\label{nuc}
\eeq
where $\expect{\phi_S}{\phi_S} =1$, $\expect{\phi_S}{\phi_D}=0$ 
and $\expect{\phi_D}{\phi_D}=1$ so that the sum of the 
amplitudes squared is unity: $|a_N|^2+|b_N|^2=1$.
The two components of the wave function are taken to 
have the following forms, respectively:
\beqa
&&\phi_S^{j_3} = \varphi_0(\P) \, \chi_{SF}^{S;j_3} \, ,\\ 
&&\phi_D^{j_3} = {1\over \sqrt{2}} \sum_{m s} 
(2 \thalf m s| \ohalf j_3)
\bigl\{
\kappa^2 Y_{2m}(\hat \kappa)\, 
\chi_F^{MS} 
 +  q^2 Y_{2m}(\hat q) \, \chi_F^{MA} \bigr\}\varphi_2(\P) 
\chi_S^{S;s}\, . 
\eeqa
The spatial $S-$ and $D-$state wave functions 
$\varphi_0(\P)$ and $\varphi_2(\P)$ are functions of the 
hyperspherical momentum variable 
$\P=\sqrt{2(\vec\kappa\,^2+\vec q\,^2)}$, where 
$\vec\kappa$ and $\vec q$ are the Jacobi momenta of 
the 3-quark system. Here the symmetric spin-flavor 
wave function $\chi_{SF}^{S;j_3}$ is defined as:
\beqa
\chi_{SF}^{S;j_3} &=& {1\over \sqrt{2}} 
\left[  \chi_F^{MS} \chi_S^{MS} 
+  \chi_F^{MA}  \chi_S^{MA} \right]\, ,
\eeqa
where $\chi_{F(S)}^{MS}$ is a mixed symmetry flavor (spin) 
wave function, and $\chi_{F(S)}^{MA}$ is a mixed 
antisymmetric flavor (spin) wave function. The function 
$\chi_S^{S;s}$ is a spin-3/2 spin-function.

The explicit expressions for the spatial wave functions 
are taken to be
\beqa
&&\varphi_0(\P)={\mathcal N}_0\bigl(1 + {\P^2\over 4 b^2}\bigr)^{-a}\, ,
\\
&&\varphi_2(\P)={\mathcal N}_2 {b^4\over \P^2}
\bigl( \varphi_0^{''}(\P)-{1\over \P}\varphi_0'(\P)\bigr)^2 \, .
\eeqa
${\mathcal N}_0$ and ${\mathcal N}_2$ are 
normalization constants and $a$ and $b$ are parameters. 
The form of the $D-$state wave function $\varphi_2(\P)$ 
is chosen so that it has the appropriate threshold 
behavior in the hyperspherical representation 
at small distances. 

The parameter choices, which lead to a satisfactory 
description of the electric form factor of the nucleon 
with Dirac quark currents without the $D-$state 
component, are $a=6, b=$ 600 MeV in instant and 
$a=9/4,b=640$ MeV in point form kinematics~\cite{bruno}. 
For front form kinematics the corresponding 
parameter choices are $a=4$ and $b=500$ MeV.

The wave function of the $\Delta(1232)$ resonance is 
taken to be formed of $S-$ and $D-$state components in 
a corresponding way as (cf. Eq.~(\ref{nuc}):
\beq
\phi_\Delta = a_\Delta \phi_S^\Delta + b_\Delta \phi_D^\Delta \, ,
\label{delta}
\eeq
where $\expect{\phi_S^\Delta}{\phi_S^\Delta}=1$, 
$\expect{\phi_S^\Delta}{\phi_D^\Delta}=0$ and 
$\expect{\phi_D^\Delta}{\phi_D^\Delta}=1$, so that
$|a_\Delta|^2+|b_\Delta|^2=1$.

The explicit expressions for the two components of 
the $\Delta(1232)$ wave function are:
\beqa
\phi_S^{\Delta;s_3} &=& \varphi_0(\P) \, \chi_S^{s_3} \, \chi_F^{3/2} \, 
,\\
\phi_D^{\Delta;j_3} &=& {1\over \sqrt{2}} 
\sum_{m s} (2 \ohalf m s| \thalf j_3)
\left\{\kappa^2 Y_{2m}(\hat \kappa)
\chi_S^{MS,s} + q^2 Y_{2m}(\hat q)  
\chi_S^{MA,s} \right\}\nonumber\\
&&\varphi_2(\P)\,\chi_F^{3/2;T}\, . 
\eeqa

\section{${\mathbf\Delta}$(1232)-N matrix elements}
\label{sec:ND}

The calculation of the matrix elements of the single 
quark current operators in the Breit frame, with account 
of the Lorentz boosts from the initial and final baryon 
rest frames and the associated Wigner rotations may be 
carried out using the formalism given in 
Refs.~\cite{bruno,bruno2}. 

The matrix element of the current can be written 
schematically as:
\beqa
\bra{\phi_\Delta^{j'_3} } {\mathcal XSF} \ket{\phi_N^{j_3}} 
&=&
 (a_\Delta \phi_S^{\Delta;j'_3} + b_\Delta \phi_D^{\Delta;j'_3})
 {\mathcal XSF}
( a_N \phi_S^{N;j_3} + b_N \phi_D^{N;j_3}) \, .
\eeqa
Here ${\mathcal X}$ is a spatial, ${\mathcal S}$ a 
spin and ${\mathcal F}$ a flavor operator.

The flavor matrix elements for the 
$\Delta(1232)^+\rightarrow N$ transition are:
\beq
\bra{\chi_F^{S}} {\mathcal F} \ket{\chi_F^{MS}} 
= {2\over 3\sqrt{2}}\, , \qquad
\bra{\chi_F^{S}} {\mathcal F} \ket{\chi_F^{MA}} = 0\, . 
\eeq
Given these the calculation reduces to the evaluation of the 
following matrix elements of spin and spatial operators:
\beqa
&&\bra{ \phi_S^{\Delta;j'_3} }
{\mathcal XSF}
\ket{ \phi_S^{N;j_3}} = {1\over 3}
\bra{ \varphi_0 \,} {\mathcal X}\ket{  \varphi_0}
\bra{ \chi_S^{j'_3}} \, 
{\mathcal S}
\ket{\chi_{S}^{MS;j_3}} \, ,\nonumber\\
&&\bra{ \phi_S^{\Delta;j'_3} }
{\mathcal XSF}
\ket{ \phi_D^{N;j_3}} = 
{1\over 3}
\sum_{m s} (2 \thalf m s| \ohalf j_3)
 \bra{\varphi_0} {\mathcal X}  \ket{O_{m}^S}
\bra{ \chi_S^{j'_3}} {\mathcal S}
\ket{\chi_S^{S;s}} \, , \nonumber \\
&&\bra{ \phi_D^{\Delta;j'_3} }
{\mathcal XSF}
\ket{ \phi_S^{N;j_3}} = 
{1\over 3\sqrt{2}}
\sum_{m' s'} (2 \ohalf m' s'| \thalf j'_3) \nonumber \\
&&
\left(
\bra{O_{m'}^S} {\mathcal X} \ket{\varphi_0}
\bra{\chi_S^{MS;s'}} {\mathcal S} \ket{\chi_S^{MS;j_3}}
+
\bra{O_{m'}^A} {\mathcal X} \ket{\varphi_0}
\bra{\chi_S^{MA;s'}} {\mathcal S} \ket{\chi_S^{MS;j_3}}
\right) \, , \nonumber \\
&&\bra{ \phi_D^{\Delta;j'_3} }
{\mathcal XSF}
\ket{ \phi_D^{N;j_3}} = 
{1\over 3\sqrt{2}}
 \sum_{m' s'} (2 \ohalf m' s'| \thalf j'_3)
 \sum_{m s} (2 \thalf m s| \ohalf j_3)
\nonumber \\
&&
\left\{ 
\bra{O_{m'}^S}{\mathcal X} \ket{ O_{m}^S} 
\bra{\chi_S^{MS;s'}}{\mathcal S}\ket{ \chi_S^{S;s}}
 +
\bra{O_{m'}^A}{\mathcal X} \ket{ O_{m}^S} 
\bra{\chi_S^{MA;s'}}{\mathcal S}\ket{ \chi_S^{S;s}}
\right\}\, .
\eeqa

\section{Numerical signatures of D-state admixtures}
\label{sec:eds}

\subsection{The elastic nucleon form factors}

\begin{figure}[t]
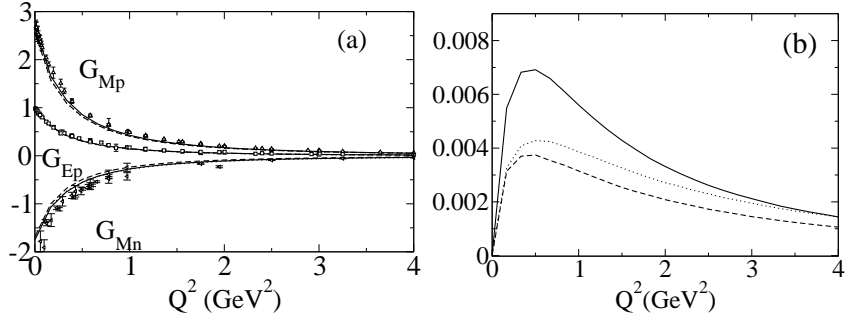

\vspace{20pt} 
\begin{center}
\mbox{\epsfig{file=fig1a, width=55mm}} 
\mbox{\epsfig{file=fig1b,width=55mm}} 
\caption{Electric and magnetic form factors 
of the proton (a) and the neutron (b) calculated 
with the $D-$state amplitude $b_N=0$ (solid), 
$b_N=0.2$ (dotted) and $b_N=-0.2$ (dashed)
in instant form kinematics. 
The experimental data are from the compilation 
of Ref.~\cite{MMD}.
\label{gemnI}}
\end{center}
\vspace{10pt} 
\end{figure}

\begin{figure}[th]
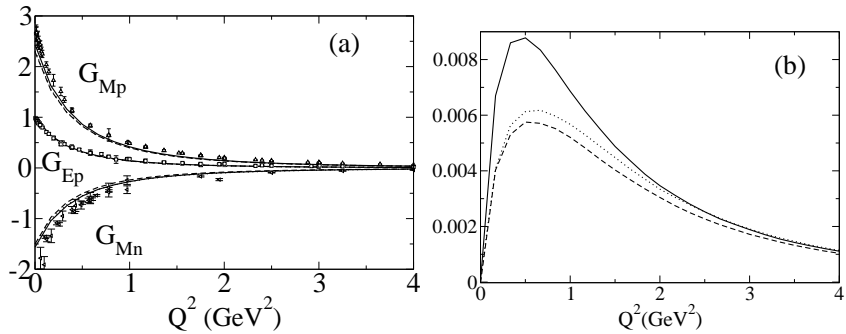

\vspace{20pt} 
\begin{center}
\mbox{\epsfig{file=fig2a, width=55mm}} 
\mbox{\epsfig{file=fig2b,width=55mm}} 
\caption{Same as Fig.~\ref{gemnI} but with 
point form kinematics.
\label{gemnP}}
\end{center}
\vspace{10pt} 
\end{figure}

\begin{figure}[th]
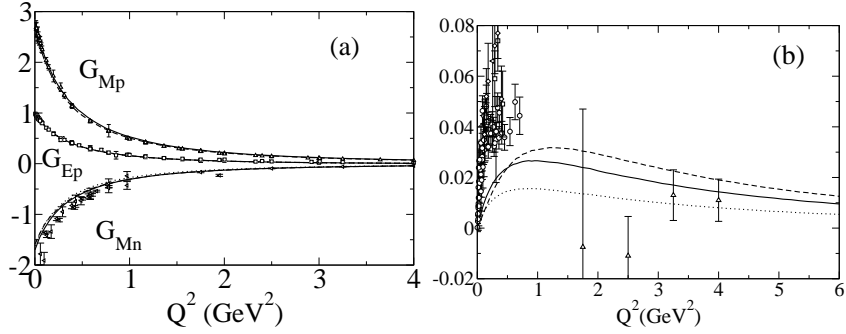

\vspace{20pt} 
\begin{center}
\mbox{\epsfig{file=fig3a, width=55mm}} 
\mbox{\epsfig{file=fig3b,width=55mm}} 
\caption{Same as  Fig.~\ref{gemnI} but for 
front form kinematics.
\label{gemnF}}
\end{center}
\vspace{20pt} 
\end{figure}

The calculated electric and magnetic form factors of
the proton and the magnetic form factor of the 
neutron as obtained in instant, point and front form 
kinematics are shown in Figs.~\ref{gemnI}(a),~\ref{gemnP}(a) 
and~\ref{gemnF}(a). The results are given with the 
$D-$state amplitude $b_N$ taken to be 0, 0.2 and 
$-$0.2 respectively. It is apparent that the effect 
of a small $D-$state component is almost negligible 
for both the magnetic form factors as well as for the 
electric form factor of the proton.

In the (b)-part of the same figures the calculated
electric form factor of the neutron is shown for 
the admixtures $b_N$. In the case of $G_{En}$ the 
description of the empirical form factor values 
deteriorates with increasing $D-$state amplitude 
for the case of instant and point forms. In the case 
of front form kinematics the calculated move towards 
the empirical values, when $b_N$ is negative. The 
effects of the $D-$state component are nonetheless 
very small as compared to that of a small admixture 
of mixed symmetry $S-$state component~\cite{bruno}.

\subsection{The $\Delta$(1232)$\rightarrow$-N$\gamma$ 
transition form factors}

\subsubsection{Instant form kinematics}

\begin{figure}[t]
\vspace{20pt} 
\begin{center}
\mbox{\epsfig{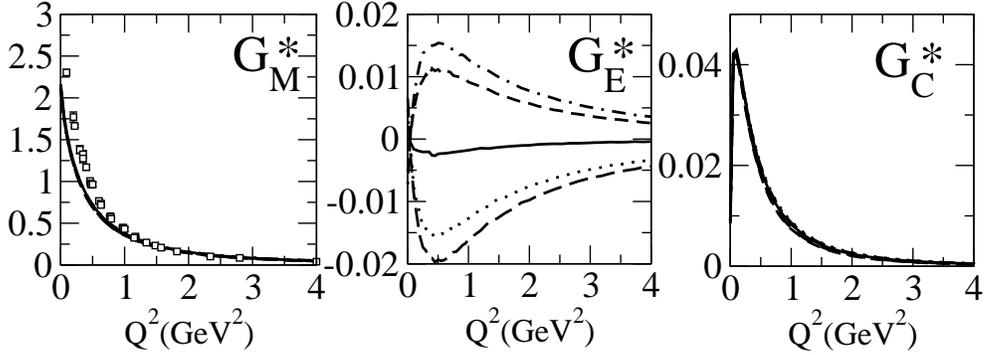}} 
\caption{$G_E^*$, $G_M^*$ and $G_C^*$ obtained in instant 
form. Solid, dotted, dashed, long-dashed and dot-dashed 
lines stand for $(b_N,b_\Delta)=$ (0,0), (0.2,0), ($-$0.2,0), 
(0,0.2) and (0,$-$0.2) respectively. In the case of $G_M^*$ 
and $G_C^*$ the lines are almost coincident. 
The data points for $G_M^*$ are from Ref.~\cite{ash}.
\label{gemci}}
\end{center}
\vspace{10pt}
\end{figure}

The three transition form factors $G_M^*$, 
$G_E^*$ and $G_C^*$ are shown as calculated 
in instant form kinematics in Fig.~\ref{gemci} for 
the following combinations of the $D-$state component 
amplitudes in the nucleon and the $\Delta(1232)$ wave 
functions: ($b_N$,$b_\Delta$)= (0,0), (0.2,0), 
($-$0.2,0), (0,0.2) and (0,$-$0.2). The results reveal 
that the presence of a $D-$state component only notably 
affects the electric transition form factor $G_E^*$. 
They also reveal that the $D-$state components in 
the nucleon and the $\Delta(1232)$ have very similar 
signatures.

\begin{figure}[h]
\vspace{35pt} 
\begin{center}
\mbox{\epsfig{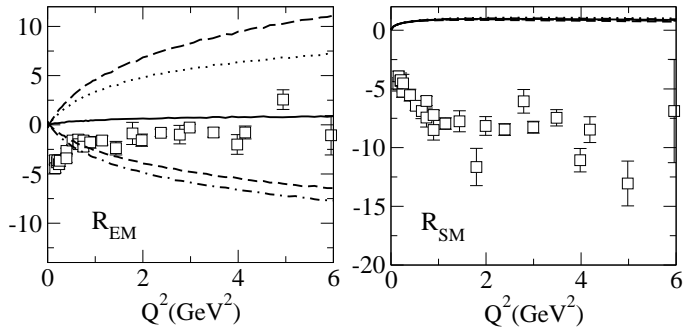}} 
\caption{$R_{EM}$ and $R_{SM}$ ratios in percent obtained 
in instant form. Solid, dotted, dashed, long-dashed and dot-dashed 
lines stand for $(b_N,b_\Delta)=$ (0,0), (0.2,0), ($-$0.2,0)
(0,0.2) and (0,$-$0.2) respectively. The experimental data 
are from the compilation of Ref.~\cite{Harry}.
\label{emratioI}}
\end{center}
\vspace{10pt}
\end{figure}

In Fig.~\ref{emratioI} the corresponding calculated 
values of the ratios $R_{EM}$ and $R_{SM}$ are shown. 
The result shows that the effect of the presence of a 
small $D-$state component is very small on the latter 
ratio. In the case of the $E2/M1$ ratio $R_{EM}$ the 
addition of the small $D-$state component does in 
contrast lead to a change in sign of the calculated 
value for the case of $b_{N(\Delta)}<0$, so that the 
calculated value takes the same sign the empirical 
values~\cite{Harry} for $Q^2 \ge 0.5$ GeV$^2$. The 
notable empirical structure at lower values of 
invariant momentum transfer is however not described 
by $D-$state configurations alone. The extant data 
are far from the asymptotic behavior $R_{EM} \to 1$ 
for $Q^2\to \infty$ suggested by perturbative 
QCD~\cite{Carlson,Latifa}.

In the case of the ratio $R_{SM}$ the quark model 
results are not brought any closer towards the 
empirical values~\cite{Harry} by introduction of 
$D-$state components.

\subsubsection{Point form kinematics}

\begin{figure}[t]
\vspace{20pt} 
\begin{center}
\mbox{\epsfig{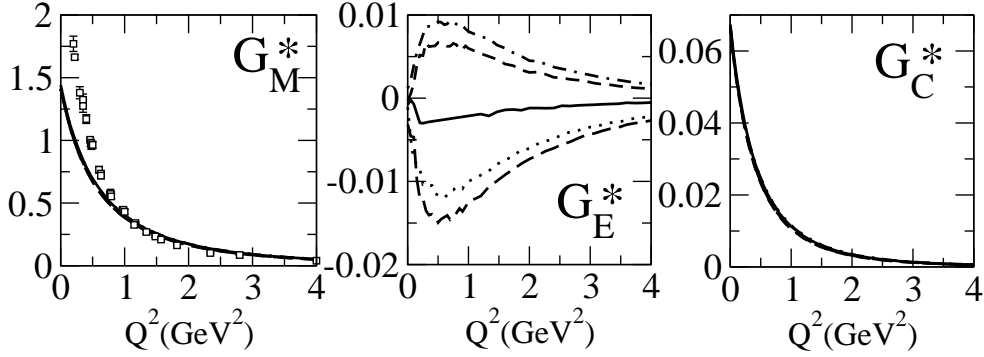}} 
\caption{$G_E^*$, $G_M^*$ and $G_C^*$ obtained in 
point form. Solid, dotted, dashed, long-dashed and dot-dashed 
lines stand for $(b_N,b_\Delta)=$ (0,0), (0.2,0), ($-$0.2,0)
(0,0.2) and (0,$-$0.2) respectively. In the case of $G_M^*$ 
and $G_C^*$ the lines are almost coincident. 
The data points for $G_M^*$ are from Ref.~\cite{ash}.
\label{gemcp}}
\end{center}
\vspace{10pt} 
\end{figure}

The transition form factors calculated in point form
with the same set of $D-$state components are shown
in Figs.~\ref{gemcp}. The results are qualitatively
similar to those obtained in instant form. The values
for the transition magnetic moment form factor do 
however fall notably below those obtained with instant 
form kinematics at low values of momentum transfer, 
and hence further from the experimental values as well.

\begin{figure}[th]
\vspace{20pt} 
\begin{center}
\mbox{\epsfig{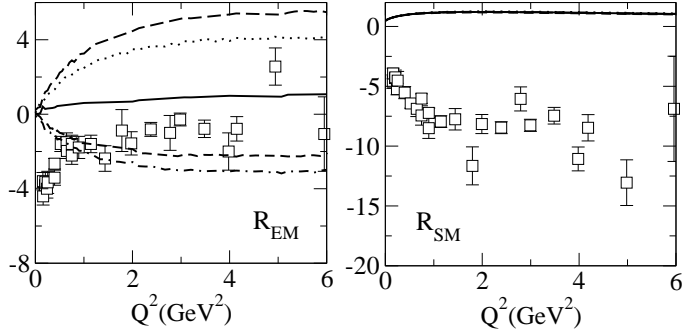}} 
\caption{$R_{EM}$ and $R_{SM}$ ratios in percent obtained 
in point form. Solid, dotted, dashed, long-dashed and dot-dashed 
lines stand for $(b_N,b_\Delta)=$ (0,0), (0.2,0), ($-$0.2,0)
(0,0.2) and (0,$-$0.2) respectively. The experimental data 
are from the compilation of Ref.~\cite{Harry}.\label{emratioP}}
\label{remp}
\end{center}
\vspace{10pt} 
\end{figure}

The values for $G_E^*$ are also correspondingly smaller,
which on the other hand leads to a better description 
of the empirical values for $R_{EM}$ with the $D-$state 
amplitudes considered. The corresponding calculated 
values of the ratios $R_{EM}$ and $R_{SM}$ are shown in
Fig.~\ref{remp}. The introduction of the $D-$state 
component does not by itself capture the low 
momentum transfer peak of $R_{EM}$ with this form
of kinematics.

In the case of point form the introduction of $D-$state 
components also have but an insignificant effect on the 
calculated values of the Coulomb form factor $G_C^*$.

\subsubsection{Front form kinematics}

\begin{figure}[t]
\vspace{20pt} 
\begin{center}
\mbox{\epsfig{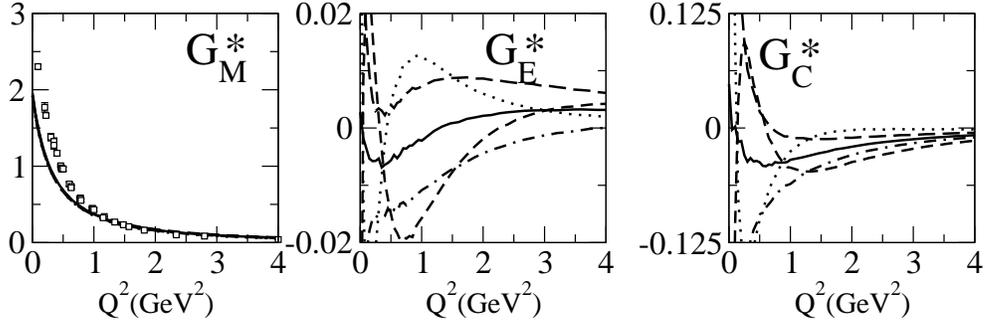}} 
\caption{$G_E^*$, $G_M^*$ and $G_C^*$ obtained in front
form. The solid, dotted, dashed, long-dashed and dot-dashed 
lines stand for $(b_N,b_\Delta)=$ (0,0), (0.2,0), ($-$0.2,0)
(0,0.2) and (0,$-$0.2) respectively. In the case of $G_M^*$ 
and $G_C^*$ the lines are almost coincident. 
The data points for $G_M^*$ are from Ref.~\cite{ash}.
\label{gemcf}}
\end{center}
\end{figure}

The transition form factors obtained in front form 
kinematics, in the reference frame where $Q^+ =0$, 
are shown in Fig.~\ref{gemcf}. These form factors 
were calculated by using only the first three 
relations in Eq.~(\ref{Qplus0}). As noted above, the
results depend on which set of spin amplitudes are 
employed~\cite{simula}.
 
The results for the magnetic transition form factor
$G_M^*$ are similar to those obtained with instant form
kinematics, and reveal very little sensitivity to the
presence of $D-$state components in the nucleon and
$\Delta(1232)$ wave function. The front form results for 
both the electric and Coulomb transition form factors
are however very sensitive to the $D-$state components,
as already noted in Ref.~\cite{simula}.

\begin{figure}[th]
\vspace{30pt} 
\begin{center}
\mbox{\epsfig{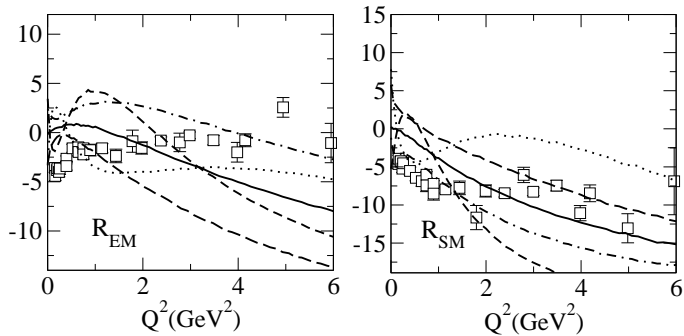}} 
\caption{$R_{EM}$ and $R_{SM}$ ratios in percent obtained 
in front form. Solid, dotted, dashed, long-dashed and dot-dashed 
lines stand for $(b_N,b_\Delta)=$ (0,0), (0.2,0), ($-$0.2,0)
(0,0.2) and (0,$-$0.2) respectively. The experimental data 
are from the compilation of Ref.~\cite{Harry}.\label{emratioF}}
\end{center}
\end{figure}

The corresponding results for the ratios $R_{EM}$ and 
$R_{SM}$ are shown in Fig.~\ref{emratioF}. The strong 
sensitivity to the $D-$state component is again evident 
from these figures. It is also found that the introduction 
of $D-$state components does not in general reduce the 
disagreement between the calculated and empirical values. 
For $R_{EM}$ the best description appears with a finite 
$D-$state component in the nucleon rather than in 
the $\Delta(1232)$ wave function.

\begin{figure}[t]
\vspace{30pt} 
\begin{center}
\mbox{\epsfig{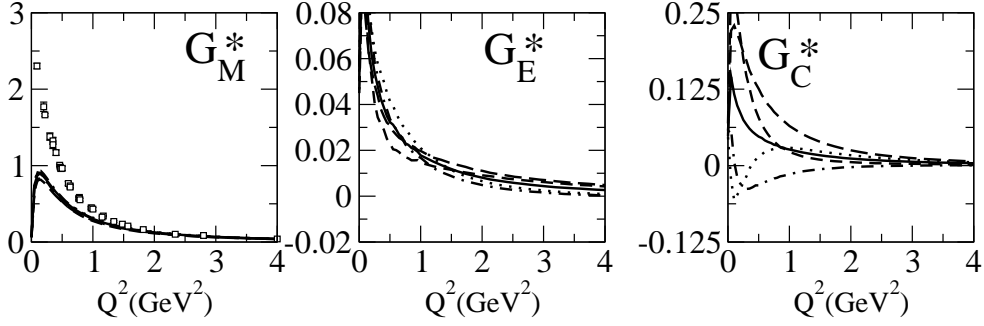}} 
\caption{$G_E^*$, $G_M^*$ and $G_C^*$ obtained in front
form in the $Q^+ \neq 0$ frame. Solid, dotted, dashed, 
long-dashed and dot-dashed lines stand for 
$(b_N,b_\Delta)=$ (0,0), (0.2,0), ($-$0.2,0)
(0,0.2) and (0,$-$0.2) respectively. In the case of $G_M^*$ 
and $G_C^*$ the lines are almost coincident. 
The data points for $G_M^*$ are from Ref.~\cite{ash}.
\label{gemcfnq}}
\end{center}
\end{figure}

For transition form factors it should a priori be more 
natural to perform the calculation in the frame 
where $Q^+\neq 0$, as this allows smooth extrapolation 
to the timelike region. The transition form factors 
calculated in front form in this frame are shown in 
Fig.~\ref{gemcfnq}. The results differ markedly from 
those obtained in the frame where $Q^+ =0$. The magnetic 
transition form factor drops to 0 at $Q^2=0$, in 
disagreement with the empirical values. The calculated 
electric and Coulomb transition form factors also 
differ considerably from those obtained in the frame 
in which $Q^+ =0$. 

The zero in $G_M^*$ at $Q^2=0$ may be traced to the form
of the relations (\ref{Qplusne0}) between the spin matrix
elements and the invariant form factors $G_j$. If these
equations are solved for the latter and the results
are inserted in (\ref{oursEMC}) it emerges that in this frame
the transition form factors are linear combinations
of the first three spin amplitudes with finite and
nonvanishing coefficients in the limit $Q^2\rightarrow 0$.
As all these spin amplitudes vanish with $Q^2$ it
follows that all three transition form factors
vanish with $Q^2$ as well.

\begin{figure}[th]
\vspace{30pt} 
\begin{center}
\mbox{\epsfig{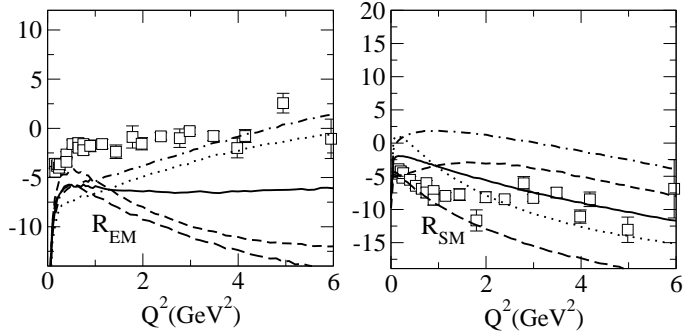}} 
\caption{$R_{EM}$ and $R_{SM}$ ratios in percent obtained 
in front form in the $Q^+ \neq 0$ frame. Solid, dotted, dashed,
 long-dashed and dot-dashed 
lines stand for $(b_N,b_\Delta)=$ (0,0), (0.2,0), ($-$0.2,0)
(0,0.2) and (0,$-$0.2) respectively. The experimental data 
are from the compilation of Ref.~\cite{Harry}.\label{emratioFnoq}}
\end{center}
\vspace{10pt} 
\end{figure}

The ratios $R_{EM}$ and $R_{SM}$  that are obtained in
front form kinematics in the frame $Q^+ \neq 0$ are shown
in Fig.{\ref{emratioFnoq}. They differ considerably
from the corresponding results obtained in the frame in
which $Q^+ =0$. The ratio $R_{SM}$ in particular has 
a shape that deviates strongly from the empirical
shape because of the small values of $G_M^*$ near
$Q^2=0$.

\section{Conclusions}

Above a comprehensive study of the effect
of $D-$state admixtures in the wave functions on 
electromagnetic form factors of the
nucleon and the
$\Delta(1232)-N$ transition form factors 
were carried out in three forms of relativistic
kinematics with single constituent currents
for pointlike quarks. The effects of the
$D-$state admixtures on the elastic 
form factors of the proton and the magnetic
form factor of the neutron were shown to be small.
Only in the case of the electric form factor
of the neutron is there a notable effect, although
this was found to be smaller than that of a
small mixed symmetry $S-$state component.

In instant and point form kinematics the effects of
the $D-$state components on the magnetic and
Coulomb
$\Delta(1232)-N$ transition form factor were 
likewise found to be but minor. A notable
sensitivity the $D-$state component was however
found in the case of the electric
$\Delta(1232)-N$ transition form factor. 
Consequently the ratio $R_{EM}$ is very
sensitive to the $D-$state, while
the ratio $R_{SM}$ is not.

In the case of front form kinematics the
impulse approximation was found to violate the
angular condition for the $\Delta(1232)-N$
spin matrix element to such a large extent that
no useful predictions could be obtained for the
electric and Coulomb transition form factors.
This problem has also been noted in Ref.~\cite{simula,simula2}.
The calculated magnetic transition form factor
is, however, similar to those obtained in front
and instant form kinematics when calculated in the
reference frame in which $Q^+ =0$. 

The empirically found structure in $R_{EM}$ at
low momentum transfer could not be described
by the inclusion of $D-$state deformation
of the nucleon and $\Delta(1232)$ 
wave functions. Since this structure, if confirmed,
represents a long range feature, and as it
is well described in the coupled channel
model of Sato and Lee~\cite{satolee}, it is
most likely due to a long range pionic
fluctuation.\\

\centerline{\bf Acknowledgment}
Instructive discussions with Drs. F. Coester and
T.-S. H. Lee are
acknowledged. Research supported in part by the 
Academy of Finland through grant 54038 and the 
European Euridice network HPRN-CT-2002-00311.


\begin{thebibliography}{30}

\bibitem{gmiller1} G. A. Miller, 
Phys. Rev. {\bf C68} (2003) 022201.

\bibitem{ralston} P. Jain and J. P. Ralston, 
hep-ph/0306194.

\bibitem{bern} A. M. Bernstein, 
Eur. Phys. J. {\bf A17} (2003) 249.

\bibitem{mainz} R. Beck, 
Phys. Rev. Lett. {\bf 78} (1997) 606.

\bibitem{satolee}  T.~Sato and T.~S.~H.~Lee,
Phys.\ Rev.\ C {\bf 54} (1996) 2660.

\bibitem{satolee2}
T.~Sato and T.~S.~H.~Lee,
Phys.\ Rev.\ C {\bf 63} (2001) 055201.

\bibitem{Dirac} P. A. M. Dirac, 
Rev. Mod. Phys. {\bf 49} (1949) 392.

\bibitem{bruno} B. Juli\'a D\'{\i}az, D. O. Riska and
F. Coester,
Phys. Rev. {\bf C69} (2004) 035212.

\bibitem{Weber} J.~Bienkowska, Z.~Dziembowski and H.~J.~Weber,
Phys.\ Rev.\ Lett.\  {\bf 59} (1987) 624
[Erratum-ibid.\  {\bf 59} (1987) 1790].

\bibitem{Coester92}
F.~Coester, Prog.\ Part.\ Nucl.\ Phys.\  {\bf 29} (1992) 1.

\bibitem{simula}
F.~Cardarelli, E.~Pace, G.~Salme and S.~Simula,
Nucl.\ Phys.\ A {\bf 623} (1997) 361C;

\bibitem{simula2}
F.~Cardarelli, E.~Pace, G.~Salme and S.~Simula,
Phys.\ Lett.\ B {\bf 371} (1996) 7; 
E.~Pace, G.~Salme, F.~Cardarelli and S.~Simula,
Nucl.\ Phys.\ A {\bf 666} (2000) 33

\bibitem{Scadron} H. F. Jones and M. D. Scadron, 
Ann. Phys. {\bf 81} (1973) 1.

\bibitem{devenish}
R.~C.~E.~Devenish, T.~S.~Eisenschitz and J.~G.~Korner,
Phys.\ Rev.\ D {\bf 14} (1976) 3063.

\bibitem{bruno2}B. Juli\'a D\'{\i}az, D. O. Riska and
F. Coester, nucl-th/0406015

\bibitem{MMD} P. Mergell, U.-G. Mei\ss ner, D. Drechsel, 
Nucl. Phys. {\bf A 596}, 367 (1996).

\bibitem{ash} W. W. Ash et al., Phys. Lett. {\bf B24} (1967) 165;
W. Bartel et al., {\it ibid,} {\bf B28} (1968) 148;
F. Foster and G. Hughes, Rep. Prog. Phys. {\bf 46} (1983) 1445;
S. Stein et al., Phys. Rev. {\bf D12} (1975) 1884;
V. V. Frolov et al., Phys. Rev. Lett. {\bf 82} (1999) 45;
S. Galster et al., Phys. Rev. {\bf D5} (1972) 519

\bibitem{Harry}
V.~D.~Burkert and T.~S.~H.~Lee, nucl-ex/0407020.

\bibitem{Carlson} C.~E.~Carlson,
Phys.\ Rev.\ D {\bf 34} (1986) 2704.

\bibitem{Latifa} V.D. Burkert and L. Elouadrhiri, 
Phys. Rev. Lett. {\bf 75} (1995) 3614.


\end{thebibliography}
\end{document}